\documentclass{article}

\usepackage{arxiv}

\usepackage[utf8]{inputenc} 
\usepackage[T1]{fontenc}    
\usepackage{hyperref}       
\usepackage{url}            
\usepackage{booktabs}       
\usepackage{amsfonts}       
\usepackage{nicefrac}       
\usepackage{microtype}      
\usepackage{lipsum}
\usepackage{amsfonts}
\usepackage{graphics}
\usepackage{amsthm}
\usepackage{amsmath}
\usepackage{graphicx}
\usepackage{amssymb}
\usepackage{pifont}
\usepackage{tikz}
\usepackage[]{algorithm2e}
\usepackage{todonotes}
\usepackage{pdflscape}
\usepackage{floatpag,mwe}
\usepackage{changepage}
\usepackage{adjustbox}
\usepackage{geometry}
\usepackage{parskip}
\usepackage{hyperref}

\DeclareMathOperator*{\argminB}{argmin}

\DeclareMathOperator*{\corr}{corr}
\DeclareMathOperator*{\dist}{dist}

\usepackage{lineno}

\title{TbCAPs: A ToolBox for Co-Activation Pattern Analysis}

\author{
  Thomas A. W. Bolton \\
  Institute of Bioengineering \\
  \'{E}cole Polytechnique F\'{e}d\'{e}rale de Lausanne (EPFL) \\
  Lausanne, Switzerland \\
  \texttt{thomas.bolton@epfl.ch} \\
   \And
 Constantin Tuleasca \\
  Faculty of Biology and Medicine \\
  Lausanne University (UNIL) \\
  Lausanne, Switzerland \\
  \And
  Gwladys Rey \\
  Department of Neuroscience \\
  University of Geneva (UNIGE) \\
  Geneva, Switzerland \\
  \And
  Diana Wotruba \\
  Program for Sustainable Development of Mental Health \\
  University of Zurich \\
  Zurich, Switzerland \\
  \And
  Julian Gaviria \\
  Department of Neuroscience \\
  University of Geneva (UNIGE) \\
  Geneva, Switzerland \\
  \And
  Herberto Dhanis \\
  Center for Neuroprosthetics \\
  \'{E}cole Polytechnique F\'{e}d\'{e}rale de Lausanne (EPFL) \\
  Lausanne, Switzerland \\
  \And
  Eva Blondiaux \\
  Center for Neuroprosthetics \\
  \'{E}cole Polytechnique F\'{e}d\'{e}rale de Lausanne (EPFL) \\
  Lausanne, Switzerland \\
  \And
  Baptiste Gauthier \\
  Center for Neuroprosthetics \\
  \'{E}cole Polytechnique F\'{e}d\'{e}rale de Lausanne (EPFL) \\
  Lausanne, Switzerland \\
  \And
  Lukasz Smigielski \\
  Center for Neuroprosthetics \\
  \'{E}cole Polytechnique F\'{e}d\'{e}rale de Lausanne (EPFL) \\
  Lausanne, Switzerland \\
  \And
  Dimitri Van De Ville \\
  Institute of Bioengineering \\
  \'{E}cole Polytechnique F\'{e}d\'{e}rale de Lausanne (EPFL) \\
  Lausanne, Switzerland \\
}

\begin{document}

\maketitle

\begin{abstract}
Functional magnetic resonance imaging provides rich spatio-temporal data of human brain activity during task and rest. Many recent efforts have focussed on characterising dynamics of brain activity. One notable instance is co-activation pattern (CAP) analysis, a frame-wise analytical approach that disentangles the different functional brain networks interacting with a user-defined seed region. While promising applications in various clinical settings have been demonstrated, there is not yet any centralised, publicly accessible resource to facilitate the deployment of the technique. 

Here, we release a working version of TbCAPs, a new toolbox for CAP analysis, which includes all steps of the analytical pipeline, introduces new methodological developments that build on already existing concepts, and enables a facilitated inspection of CAPs and resulting metrics of brain dynamics. The toolbox is available on a public academic repository \url{https://c4science.ch/source/CAP_Toolbox.git}.

In addition, to illustrate the feasibility and usefulness of our pipeline, we describe an application to the study of human cognition. CAPs are constructed from resting-state fMRI using as seed the right dorsolateral prefrontal cortex, and, in a separate sample, we successfully predict a behavioural measure of continuous attentional performance from the metrics of CAP dynamics (R=0.59).
 \end{abstract}


\section{Introduction}

Functional magnetic resonance imaging (fMRI) has enabled to track temporal changes in activity levels at the whole-brain scale by means of the blood oxygenation level-dependent (BOLD) contrast, a proxy for neural activation~\cite{Logothetis2001}. In addition to more traditional task-based studies in which BOLD changes are mapped to a paradigm of interest~\cite{Friston1994}, the characterisation of statistical interdependence between remote brain locations (termed \textit{functional connectivity}~\cite{Friston1994b}) in the resting-state, and the concomitant definition of large-scale \textit{resting-state brain networks} (RSNs), has been a popular endeavour~\cite{Biswal1995,Fox2005,Damoiseaux2006,Power2011}, with great benefits for the understanding of cognition and disease~\cite{VanDenHeuvel2010,Greicius2008,Fox2010}.

Over the past years, it has become increasingly appreciated that cross-regional relationships do not remain static over the course of a full scanning session~\cite{Chang2010}: instead, a given region rearranges its interactions along time, in ways that have been addressed with very diverse analytical tools (see~\cite{Hutchison2013b,Preti2017} for exhaustive reviews of the \textit{dynamic functional connectivity} field).

In one family of approaches that has been developed, it is assumed that only few salient time points contain the information of interest that shapes whole-brain correlational relationships; selecting only these frames, by means of a seed-based thresholding process, already enables to derive accurate RSN maps, even if as little as 10\% of data points is retained~\cite{Tagliazucchi2012b}. The analysis then moves from a second-order correlation-based characterisation to a first-order activation viewpoint, and reduces computational load, a desirable feat in light of the numerous large-scale acquisition initiatives embraced by the fMRI community~\cite{VanEssen2013,Nooner2012,Holmes2015}.

Building on this point-process analysis concept, and inspired by the dynamic viewpoint on resting-state brain function,~\cite{Liu2013} hypothesised that at different moments in time, the seed region of interest would display distinct interactions with the rest of the brain. A k-means clustering step was thus appended to frame selection, so that fMRI volumes with a large enough seed activity would be partitioned into a limited set of \textit{co-activation patterns} (CAPs).

Since then, co-activation pattern analysis has started to gain momentum as a potent tool to reveal functional brain dynamics subtleties: analyses taking the posterior cingulate cortex (PCC) as a seed revealed alterations of spatial intensity level and occurrence in specific CAPs~\cite{Amico2014,DiPerri2018}, while in adolescent depression,~\cite{Kaiser2019} showed that the time spent in a specific frontoinsular-default network CAP positively correlated with symptoms severity. CAP analysis also enabled to track the renormalisation of CAP occurrences in patients with essential tremor following surgical intervention~\cite{Tuleasca2019}.

In parallel to clinical applications, the technical details of the approach have also been addressed, in terms of retaining activation versus deactivation time points~\cite{Di2013}, extending the approach to the whole brain~\cite{Liu2013b}, designing novel metrics of interest~\cite{Chen2015}, or constraining the extent of spatial overlap across CAPs~\cite{Zhuang2018}. For more details, the reader is pointed at the recent review of~\cite{Liu2018}.

Here, we wish to further foster the development of CAP analysis by releasing a dedicated toolbox, which enables to easily navigate through the steps of the analytical pipeline through a graphical user interface, and also offers additional technical developments regarding frame selection and metrics computation. While the mathematical underpinnings of CAP analysis are relatively straightforward, we hope that providing such a resource will encourage practitioners to embrace the method, and it will become easier to compare CAP analyses based on subtle, but sometimes important, differences in the processing pipeline.

In addition, to exemplify the use of our toolbox, we describe an application of CAP analysis in the yet unaddressed setting of predicting cognitive skills: in a battery of healthy individuals, we show that continuous performance in a visual attention and vigilance task correlates with the expression profile of task-positive network (TPN) CAPs.

\section{Materials and Methods}

\subsection{Co-activation pattern analysis theory}

Let us consider the data matrix $\mathbf{X}_s\in \mathbb{R}^{V \times T}$ for subject $s$, where $V$ is the number of voxels to consider in the analysis and $T$ the number of time points. Each voxel-wise time course is temporally z-scored, so that $\mu_i=\frac{\sum_{t=1}^{T}X_{s}(i,t)}{T}=0$ and $\sigma_i=\sqrt{\frac{\sum_{t=1}^{T}(X_{s}(i,t)-\mu_i)^2}{T-1}}=1$, for all $i=1,2,\cdots,V$.

Co-activation pattern analysis requires the definition of a seed region, whose interactions with the rest of the brain will be probed. Formally, a set of voxels $\mathcal{S}$ that one wishes to consider is specified, and a time point $t$ of the seed activation time course is then given by:
\begin{equation*}
S_\text{seed}(t) = \frac{\sum_{i\in\mathcal{S}} X_{s}(i,t)}{|\mathcal{S}|},\quad\text{for all}\quad t\in 1,2,\cdots,T.
\end{equation*}

Only time points when the seed time course takes sufficiently extreme values (denoting significant seed (de)activation) are considered. Let the activation threshold be $T$, we then construct the set $\mathcal{T}_s$ of time points that satisfy $S_\text{seed}(t)>T$ (if we wish to consider solely activation moments) or $S_\text{seed}(t)<-T$ (if we are interested in seed deactivation time points).

In this work, in addition to the above, we propose an extension in which more than one seed region can be considered: for each seed $j$, a set of time points $\mathcal{T}_{s,j}$ is derived. Assuming $J$ separate seeds, one can then consider the time points when all seed time courses jointly take extreme values:
\begin{equation*}
\mathcal{T}_{s,\text{Intersection}} = \bigcap\limits^{J}_{j=1}\mathcal{T}_{s,j}.
\end{equation*} 

Alternatively, one may instead be interested in the moments when at least one of the seed regions becomes strongly (de)active:
\begin{equation*}
\mathcal{T}_{s,\text{Union}} = \bigcup\limits^{J}_{j=1}\mathcal{T}_{s,j}.
\end{equation*}

Finally, other additional criteria can be incorporated at the time point selection step: for instance, given the deleterious impact that head motion exerts on BOLD signals even following standard preprocessing~\cite{Power2012,VanDijk2012,Satterthwaite2012}, it may be desirable to only retain the frames for which framewise displacement does not exceed a threshold $\text{FD}_\text{limit}$.

After having selected the frames to keep for each subject, the next step is the population-level clustering of data points into CAPs. K-means clustering is used for this purpose, to optimise:
\begin{equation}
\argminB_{\mathbf{\mathcal{C}}} \sum_{k=1}^{K}\sum_{s=1}^{S}\sum_{t\in\mathcal{T}_{s}\cup\mathcal{C}_k} \text{dist}(\mathbf{X}_{s}(\cdot,t),\mathbf{c}_{k}),
\label{CAPs_EQ1}
\end{equation} 
where $K$ is the number of co-activation patterns to derive, $\mathbf{\mathcal{C}}=\{\mathcal{C}_1,\cdots,\mathcal{C}_K\}$ summarises the hard assignment of the frames to each CAP, and $\mathbf{c}_k$ is the spatial map for co-activation pattern $k$. The $\dist$ function depends on the type of distance to use in the algorithm. In addition, since k-means clustering is an iterative process with no guaranteed convergence towards the global optimum, the algorithm is run $n_\text{rep}$ times.

In several previous works using CAPs, it was also suggested to solve Equation~\ref{CAPs_EQ1} after setting to 0 the voxel intensity values that, for each frame of interest, would not be part of the largest $P_\text{P}$ or $P_\text{N}$ percents (for positive-valued and negative-valued voxels, respectively)~\cite{Liu2013,Liu2013b}.

Table~\ref{CHAPTER5_TABLE0} summarises the different parameters that are defined for CAP analysis, and also highlights the default values that we used in this work.

\begin{table}[h!]
\centering
\scriptsize
\begin{tabular}{| c | l | c |}
\hline 
Parameter & Description & Default value \\ 
\hline \hline
$J$ & Number of seeds to use & $1$ \\ 
$\mathcal{S}$ & Voxel set to use as seed & Right dorsolateral prefrontal cortex \\ 
Polarity & Sign of the seed excursions to consider & Activation \\ 
Seed combination & Whether all or at least one seed should be (de)active to retain a time point & n.a. \\ 
 $T$ & Threshold for frame selection & $1.5$ \\ 
 $\text{FD}_\text{limit}$ & Threshold of framewise displacement above which to scrub & $0.3$ mm  \\ 
 $K$ & Number of clusters to use & 16 \\ 
 $n_\text{rep}$ & Number of replicates of k-means & $50$  \\ 
 $P_\text{P}$ & Percentage of positive-valued voxels to keep in each frame for clustering & $100$ \\ 
 $P_\text{N}$ & Percentage of negative-valued voxels to keep in each frame for clustering & $100$ \\ 
 $\dist$ & Distance measure used for clustering & $\corr$ \\ \hline 
\end{tabular}
\normalsize
\caption{\textbf{Parameters to define for co-activation pattern analysis.} The seed region was extracted from a TPN independent component map derived in~\cite{Shirer2012}. Investigated cluster number values $\{K\}$ were determined through consensus clustering, a subsampling-based assessment of clustering robustness~\cite{Monti2003}, and the final choice for the analyses was defined based on an exploratory assessment of the brain/behaviour correlation significance.}
\label{CHAPTER5_TABLE0}
\end{table}

\subsection{Metrics characterising CAPs dynamics}

Once all retained frames have been assigned to CAPs representatives, it becomes possible to construct, for each subject, an empirical transition probability matrix $\mathbf{A}_s$ that summarises the likelihood to transit from a given CAP at time $t$ to another at time $t+1$. Another available piece of information regards the likelihood to transit from and back to the baseline state (when the seed was not significantly (de)active). Further, if separate subject populations are used in computing CAPs and deriving associated metrics (as in our example applications below), there are also occurrences of entries into an extra state associated to frames that could not be matched to any CAP with sufficient certainty.

\begin{figure}[h!]
\centering
\includegraphics[width=1.0\textwidth]{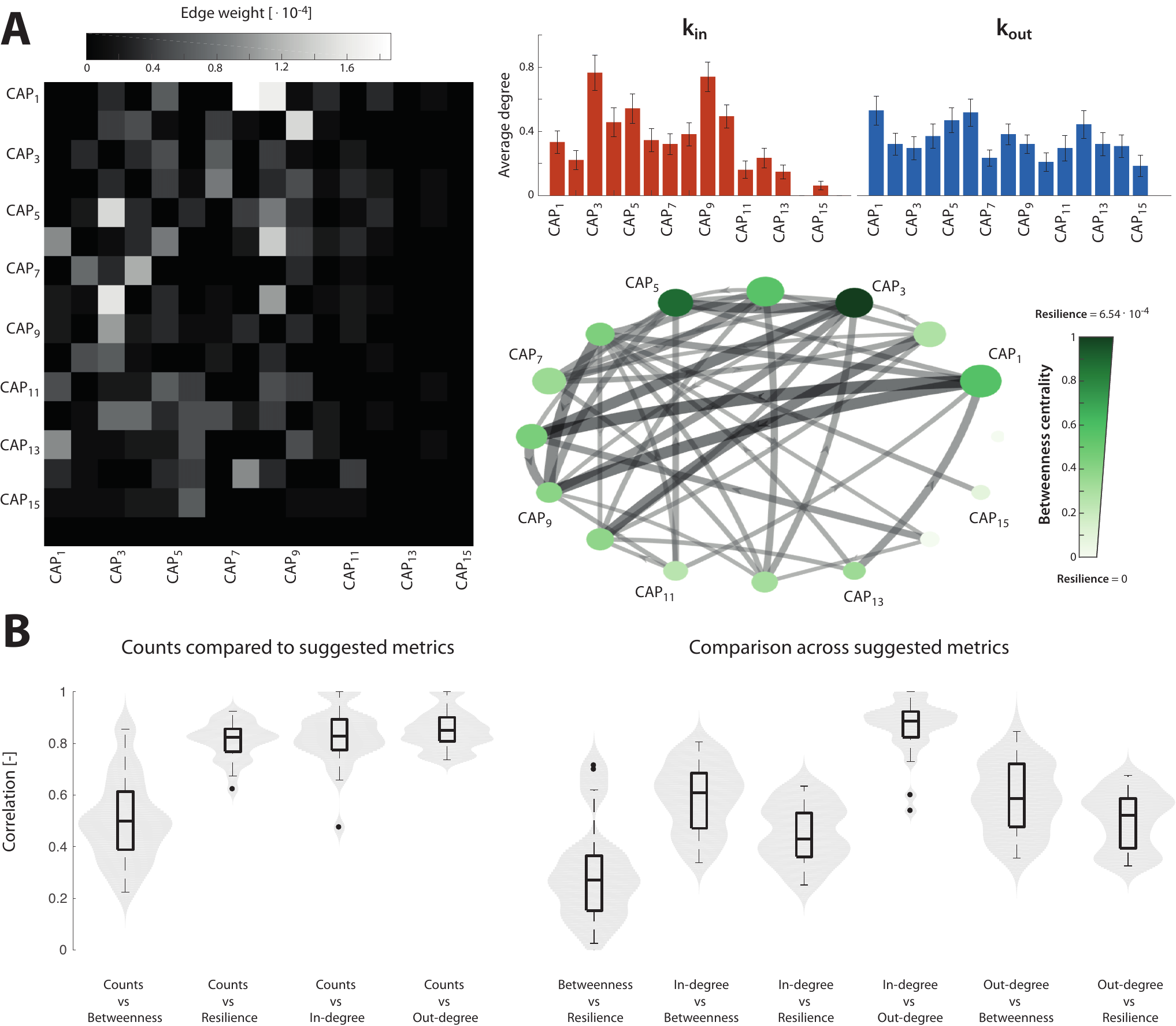}
\caption{\textbf{Generation of CAP metrics.} \textbf{(A)} Transitions across CAPs can be viewed in terms of individual transition probabilities for a total of $K^2$ features (left), or alternatively, a directional graph representation can be constructed (right) to extract in-degree (red bars), out-degree (blue bars), betweenness centrality (green colour coding of the nodes in the bottom plot) and resilience (size of the nodes) information, for a reduced total of $4K$ features. Error bars denote standard error of the mean, and the displayed transition probabilities, betweenness centrality values and resilience values are averages across subjects. \textbf{(B)} The extracted metrics contain equivalent information compared to the more traditional occurrences, and each such metric characterises partly different aspects of CAP dynamics, as seen by moderate correlations. Each box plot/violin plot representation depicts correlation values across $K$ CAPs. Illustrations from this figure are generated from the data presented in our example application, with $K=16$, $T_\text{P}=0$\%, and without assigning scrubbed volumes.}
\label{CHAPTER5_P1F1}
\end{figure}

An indicative example of averaged transition probability matrix across subjects is displayed in Figure~\ref{CHAPTER5_P1F1}A (left column). Individual elements of the transition probability matrix may be considered as such~\cite{Chen2015}, which would amount to a total of $K^2$ values per subject. 
To meaningfully lower the amount of features of interest, we propose to rather view the available information as a directional graph representation, from which a series of summarising metrics can be derived~\cite{Rubinov2010}. First, by sampling the diagonal elements of the matrix, we obtain a measure of \textit{resilience} for each CAP: that is, the likelihood to remain in the same configuration from time $t$ to $t+1$. Second, after having set the diagonal elements of the matrix to 0, we can define the \textit{in-degree} $k_\text{in}$ (how likely a CAP is visited from any other), the \textit{out-degree} $k_\text{out}$ (how likely a CAP is exited towards any other), and the \textit{betweenness centrality} (how important a CAP is regarding the shortest paths between other pairs)~\cite{Freeman1979}. In total, the feature space has thus been reduced from $K^2$ to $4K$. This alternative viewpoint is exemplified in Figure~\ref{CHAPTER5_P1F1}A (right column). 

In several works, \textit{counts} or \textit{occurrences} (that is, how many times a given CAP is expressed) were used as metrics of interest~\cite{DiPerri2018,Kaiser2019,Tuleasca2019}. We verified that our suggested metrics also include the information rendered by the counts: as seen in Figure~\ref{CHAPTER5_P1F1}B, the average correlation across CAPs between counts and in-degree, out-degree or resilience exceeded 0.8 (respectively $\rho=0.83\pm0.11$, $\rho=0.85\pm0.08$ and $\rho=0.81\pm0.07$). From pair-wise comparisons between our four metrics, it can also be seen that in-degree and out-degree are strongly correlated ($\rho=0.87\pm0.1$), while resilience and betweenness centrality capture separate information given their more moderate correlations (for resilience: $\rho=0.45\pm0.11$, $\rho=0.5\pm0.11$ and $\rho=0.3\pm0.19$ with in-degree, out-degree and betweenness centrality, respectively; for betweenness centrality: $\rho=0.59\pm0.13$ and $\rho=0.59\pm0.13$ with in-degree and out-degree, respectively). Despite their overall similarity, we decided to retain both in-degree and out-degree as they still yielded different values in specific CAP cases.

In addition to the above metrics that summarise the transitory behaviour across different CAPs, an interesting complement is the assessment of which CAPs are entered from the baseline state of seed activity, as well as of which CAPs are the ones expressed just before a return to baseline activity. With this additional information, a total of $6K$ features of interest per subject is available. These are the summarising measures that we use in our example application.

\subsection{TbCAPs: implementation}
\begin{figure}[h!]
\centering
\includegraphics[width=1.0\textwidth]{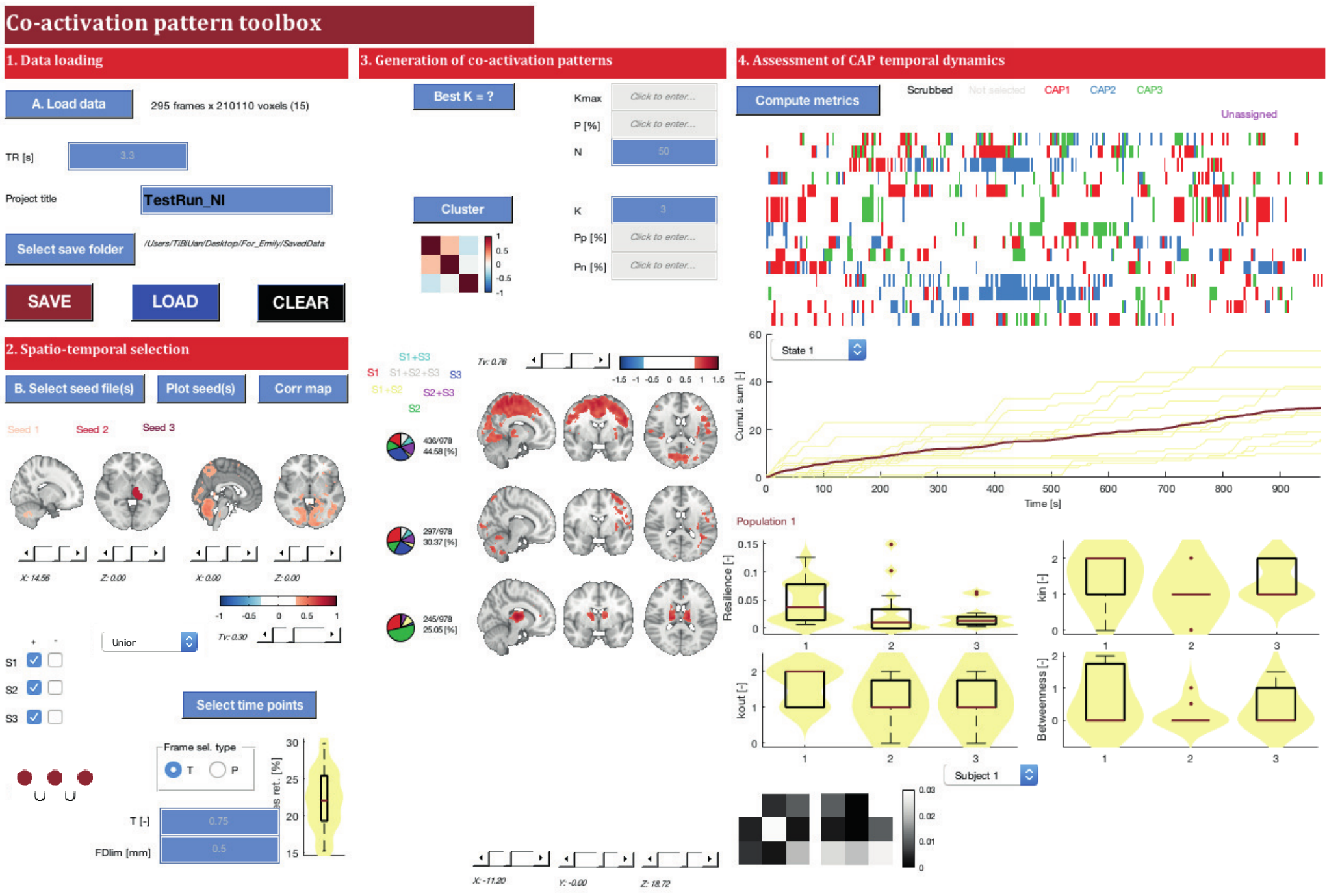}
\caption{\textbf{Illustration of the TbCAPs graphical user interface.} The typical output of a CAP analysis is displayed for a set of healthy volunteers, with three seeds chosen within a functional circuitry associated to essential tremor (this data was explicitly analysed in~\cite{Tuleasca2019}).}
\label{CHAPTER5_P1F2}
\end{figure}

We implemented the CAPs processing pipeline as a toolbox in Matlab version 2017a (The MathWorks, Natick, USA). This software is freely accessible at \url{https://c4science.ch/source/CAP_Toolbox.git}. It contains a graphical user interface to facilitate the different steps of the pipeline. In addition, we also provide a scripted version of a typical analysis pipeline for power-users. An illustrative display of the graphical user interface at the end of a typical analysis is provided in Figure~\ref{CHAPTER5_P1F2}. Next, we describe the steps to be performed by the user, and the available options at each stage of the analysis.

\subsubsection{Data preparation}

The toolbox requires four different input files, to be selected at once when clicking on the \textbf{A. Load data} button: first, the preprocessed fMRI data should be provided as a Matlab structure of size $1 \times S$. Each cell should contain a matrix of size $T \times V$. Second, framewise displacement over time for all subjects should be provided as a matrix of size $T \times S$. Third, a NIFTI file header containing the dimensionality information about the data at hand should be included. Fourth, a mask of size $V_{all}$ (all the acquired voxels) should be passed as a logical vector, setting to $1$ the $V$ voxels to include for CAP analysis.

\subsubsection{Spatio-temporal selection}

Following data loading, the user is prompted to select one or more seeds to use in the analysis: all seed files should be entered at once, each as a logical vector of size $V \times 1$. The brain area covered by the seeds can be inspected in brain slice representations, which can be navigated through by means of dedicated sliders. For now, we allow up to three separate seeds to be entered for the analyses. In addition, the interested user can also plot the seed-based correlation map associated to the first selected seed.

The next step is to select which types of events should be retained (activation versus deactivation), and if more than one seed was selected, whether all seeds should show an extreme event at once to select a time point (\textbf{Intersection} option), or if a frame should be kept as long as at least one does so (\textbf{Union} option). The user is also prompted to determine the threshold $T$ to use in selecting frames (or alternatively, a percentage of most (de)active frames to retain), and the threshold $\text{FD}_\text{limit}$ above which frames will be deemed excessively corrupted by head motion, and scrubbed out.

\subsubsection{Generation of co-activation patterns}

Regarding the subsequent generation of CAPs, if the optimal number of clusters $K$ to select is not known a priori, we offer the possibility to run consensus clustering~\cite{Monti2003}, where clustering is run many times from $K=2$ to a user-specified $K_\text{max}$ using a subsample of the data (the percentage of data points to use is specified by $P$, and the number of iterations by $N$). A good clustering solution is one for which across folds, two frames are either always clustered together, or never clustered together (but not an intermediate case). We quantify this by the Percentage of Ambiguously Clustered pairs (PAC)~\cite{Senbabaoglu2014}, and display the stability measure $1-\text{PAC}$.

When k-means clustering is run by clicking the \textbf{Cluster} button, the first $5$ CAPs with most occurrences across the subject population are displayed, and can be visually inspected. If using the \textbf{Intersection} option in a multi-seed analysis, the user is also shown pie charts summarising, for each CAP, what fraction of frames was selected in a given seed combination configuration.

\subsubsection{Metrics generation and frame assignment}

Finally, upon clicking the \textbf{Compute metrics} button, displays of CAP expression time courses, cumulative CAP expression along time, and violin plots summarising the distribution of computed metrics across subjects for each CAP, are provided. Transition probability matrices can also be inspected for each subject.

In some settings, the user may wish to compare different subject populations: this can be done by sequentially loading up to $4$ different populations at the start of the analysis. CAPs will be derived from the first population, and there is then the option to assign the frames from the other populations to the CAPs by a matching process. In doing so, the spatial correlation between a frame and the CAP to which it is most similar is compared to the distribution of spatial correlations of the frames from population 1 that belong to the CAP in question: if the $T_\text{P}$\textsuperscript{th} percentile of this distribution is exceeded, assignment is performed; else, the frame is left unassigned and belongs to an extra $(K+1)$\textsuperscript{th} cluster.

\subsection{Application to experimental fMRI data}

\subsubsection{Functional data preprocessing}

As a proof of feasibility and application of TbCAPs, we considered a sample of $181$ subjects from the \textit{Human Connectome Project}~\cite{VanEssen2013}, selected because they had at least one fully exploitable resting-state scanning session on which to apply the method, and less than 5\% of recorded behavioural entries that were missing. Details regarding acquisition parameters can be found elsewhere~\cite{Smith2013}, but briefly, the data was acquired at a TR of 0.72 s over 15 minutes (for a total of 1200 fMRI volumes), with a spatial resolution following initial preprocessing steps of 2 mm $\times$ 2 mm $\times$ 2 mm.

We started from the \textit{minimally preprocessed} resting-state data (first session, LR acquisition direction). The first 10 samples of the data were discarded. We then performed linear detrending, and regressed out low-frequency components of the discrete cosine transform basis with a cutoff frequency at 0.01 Hz. Due to collinearity with this basis, we did not regress out average white matter or cerebrospinal fluid time courses. We also chose not to regress motion parameter time courses, as motion is handled within the co-activation pattern pipeline by scrubbing, and because recent evidence points at the fact that motion regression schemes may not always be beneficial in the context of brain/behaviour analyses~\cite{Bolton2019c}. As for global signal regression, given the lack of a clear consensus~\cite{Murphy2017}, we preferred to leave the data as untouched as possible and did not include it.

Following the regression step, the data was scrubbed at a framewise displacement threshold of $0.3$ mm, and excised volumes were estimated with cubic spline interpolation. Although scrubbing is performed within TbCAPs, we reasoned that if we wished to try and assign scrubbed frames to CAPs in our additional analyses regarding head motion, it would make more sense to have previously corrected these volumes to the best of our abilities.

Then, individual fMRI volumes were smoothed at a full-width at half maximum value of 5 mm, and in order to make the analyses computationally more affordable, spatial resolution was downsampled at 3 mm $\times$ 3 mm $\times$ 3 mm. Finally, each voxel-wise time course was z-scored to obtain the resting-state functional imaging input data to our pipeline. 

\subsubsection{Selection of seed and behaviour of interest}

As a behaviour of interest to study, we selected the Short Penn Continuous Performance Test (SCPT), which quantifies continuous sustained attention~\cite{Gur2010}. In more details, participants see vertical and horizontal red lines flash on screen, and from block to block, must respond either when the lines form a number, or a letter. The lines are displayed for 300 ms, followed by a 700 ms inter-trial interval.

We started from raw behavioural entries provided by the HCP, for $951$ different subjects. There are $8$ available SCPT measures: amount of true positives, false positives, true negatives or false negatives, median response time for true positive responses, sensitivity, specificity and longest run of non-responses. In order to reduce this information into one summary measure while filling in missing behavioural entries, we performed probabilistic PCA~\cite{Bishop1999}. The output composite score positively correlated with true positives, true negatives, sensitivity and specificity ($\rho=$0.24, 1.00, 0.25 and 1.00, respectively), thus summarising overall task performance. We z-scored this output measure across subjects, in order to quantify performance with respect to the overall population. We then extracted the behavioural data related to the $181$ subjects considered in this work.

To study sustained attention, we focussed on a right dorsolateral prefrontal cortex seed from the task-positive network, which we extracted from the associated independent component map provided by~\cite{Shirer2012}. Our hypothesis was that the expression of different TPN configurations would relate to sustained attention performance.

\subsubsection{Co-activation pattern analysis details}

We resorted to a threshold $T=1.5$ to select active frames, and performed scrubbing with a framewise displacement threshold $\text{FD}_\text{limit}=0.3$ mm.

To avoid double dipping~\cite{Kriegeskorte2009}, CAPs were extracted from a randomly selected subset of $100$ subjects, while we performed correlations with behaviour for the remaining $81$ only. To determine the optimal number of clusters, we used consensus clustering~\cite{Monti2003}. We then ran k-means $n_\text{rep}=50$ separate times, keeping the best solution. We included all voxels in the analyses ($P_\text{P}=P_\text{N}=100$\%), and used spatial correlation as our distance measure; given two similarly-sized vectors $\mathbf{x}$ and $\mathbf{y}$, this thus yields $\dist(\mathbf{x},\mathbf{y})=1-\corr(\mathbf{x},\mathbf{y})$.

Following the extraction of the CAPs on our $100$ \textit{training subjects}, we determined which CAP was expressed at each retained fMRI volumes of the other $81$ subjects. To do so, we used the aforementioned assignment process with $T_\text{P}$ ranging from 0 to 100\%.

\subsubsection{Assessment of brain/behaviour relationships}

As imaging metrics of interest, we considered in-degree, out-degree, betweenness centrality and resilience for each CAP, and also included the amount of excursions from the baseline state, and the amount of excursions back to the baseline state. Thus, we generated a total of $6K$ imaging features per subject.

After having obtained the behavioral scores $\mathbf{b}\in\mathbb{R}^{81\times 1}$ and metrics ${M}\in\mathbb{R}^{81\times 6K}$ for our population of subjects, we used Partial Least Squares (PLS) analysis~\cite{McIntosh2004,Krishnan2011} to probe the existence of a brain/behaviour relationship. 

Briefly, consider a matrix of behavioural features $\mathbf{B}\in\mathbb{R}^{S\times n_\text{B}}$ and a matrix of imaging metrics $\mathbf{M}\in\mathbb{R}^{S\times n_\text{M}}$. Assuming that $n_\text{B} < n_\text{M}$, and using the singular value decomposition, the covariance between these two sets is given by:
\begin{equation*}
\mathbf{R} =  \mathbf{M}^{\top} \mathbf{B} = \mathbf{U} \mathbf{\Sigma} \mathbf{V}^{\top} = \sum_{i=1}^{n_\text{B}} \sigma_i \mathbf{u_{i}} \mathbf{v_{i}}^{\top},
\end{equation*}
where each column in $\mathbf{U}$ and $\mathbf{V}$ contains the weights (so called \textit{saliences}) that respectively multiply imaging and behavioural markers to yield a maximised covariance between both sets. The associated singular value $\sigma_i$ is proportional to the fraction of covariance explained by the component at hand. 

In our case, since $n_\text{B}=1$ (we only consider one behavioural measure), only one covariance component is retrieved, which implies $v_{1}=1$. The interesting information lies in $\mathbf{u}_{1}$: positive-valued saliences highlight metrics that are larger in subjects who show a greater cognitive ability, and negative-valued saliences are associated to metrics that, when larger, impede attentional performance.

Prior to running the algorithm, each of the $6K$ features was z-scored across subjects. In order to assess significance, we reran PLS $1000$ times after having randomly shuffled the subject entries in one of the two matrices; to non-parametrically derive a p-value, the singular value of the actual covariance component was compared to the null distribution constructed from this permutation process.

Further, to establish the stability of the salience weights, we reran PLS $1000$ times using a subsample of 80\% of the data each time, and computed a bootstrap score for each salience weight as its mean across folds divided by its standard deviation.

\subsubsection{Influence of head motion on quantified metrics}

While scrubbing enables to minimise the deleterious impacts of motion on the analysis and compute clean CAPs, discarding frames also has the potential to distort transition probability estimates. For example, a succession of three frames in the same state (which would amount to a higher resilience for the CAP in question) would not be captured if the middle frame is scrubbed out.

To verify that our findings were minimally sensitive to this effect, we ran another series of analyses in which we also performed the aforementioned assignment process on scrubbed frames, with a similar $T_\text{P}$ range as for assigning test subject frames. This way, frames strongly distorted by head motion still do not enter the analysis, but more mildly affected fMRI volumes can be matched to their CAP. We verified the reproducibility of our findings upon this additional analytical step.

\section{Results}

\begin{figure}[h!]
\centering
\includegraphics[width=1.0\textwidth]{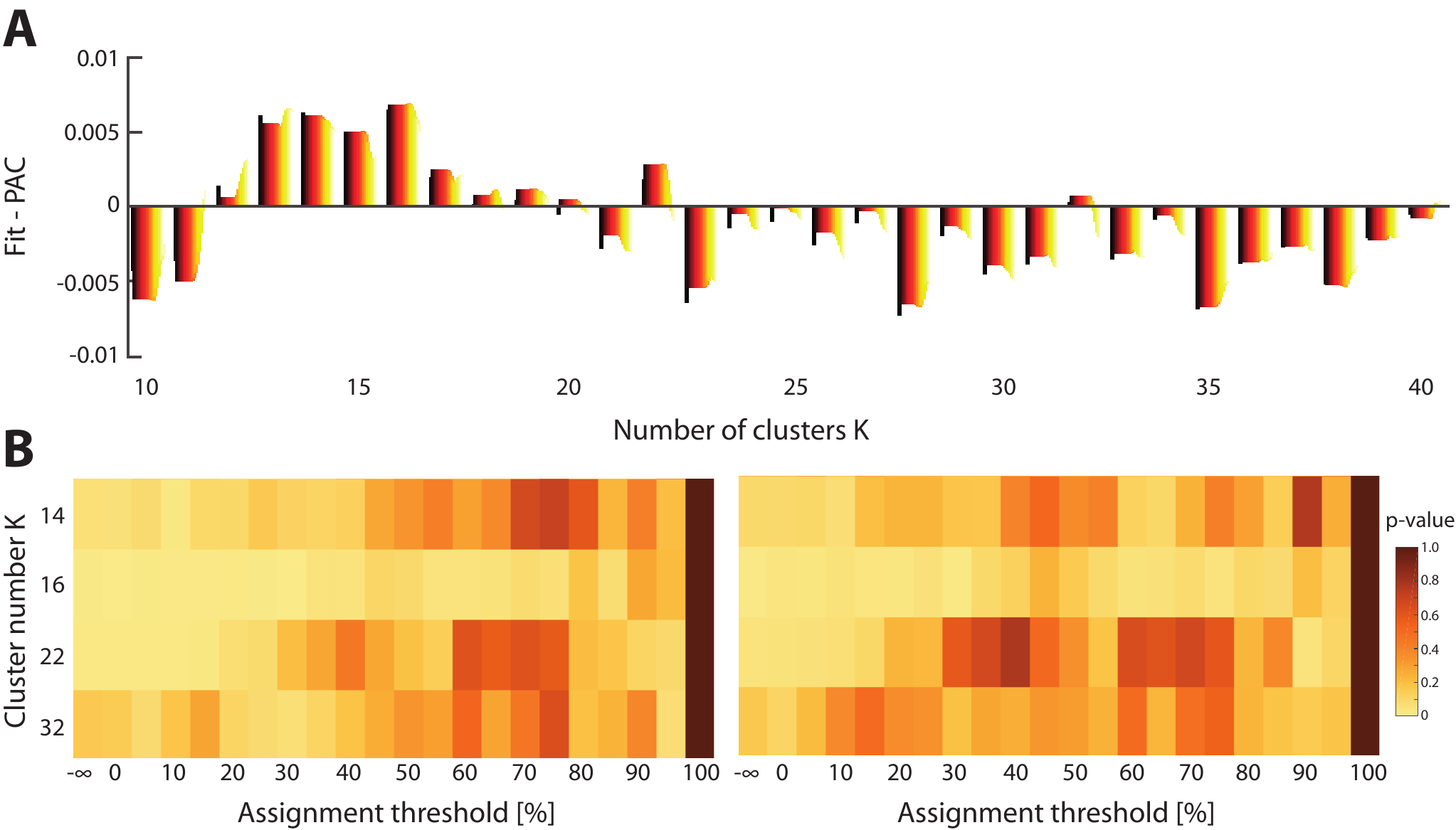}
\caption{\textbf{Parameter selection.} \textbf{(A)} Consensus clustering results for a range of K values from 10 to 40. Outputs from the algorithm are percentages of ambiguously clustered pairs (PAC); since this measure requires the definition of an interval of consensus values within which clustering is deemed \textit{ambiguous}, we investigated a range of values as colour-coded from black to yellow. For each value, we fitted a decreasing exponential to capture the overall tendency of PAC values going down with larger cluster number. The y-axis of the plot depicts the difference between this fitted value and the actual value: thus, a positive difference means that the considered cluster number is more satisfying than what would be predicted in terms of the overall behaviour. \textbf{(B)} Across candidate cluster number values selected from consensus clustering, and assignment threshold values, significance of the relationship between CAP metrics and attentional abilities, as quantified with the p-value obtained by PLS analysis. The left panel depicts the results for which scrubbed frames are not considered at all, while in the right panel, scrubbed frames were also assigned to the CAPs. The infinity symbol is used to depict a case in which assignment is done for all frames, even if a frame is not sufficiently close to any CAP when comparing its spatial correlation to the associated correlation distribution.}
\label{CHAPTER5_P1F3}
\end{figure}

Consensus clustering results are displayed in Figure~\ref{CHAPTER5_P1F3}A for $K$ values ranging from 10 to 40. Positive peaks highlight good candidate values (see figure legend for details); such values are present for diverse numbers of clusters (more notably at $K=16,22,32$). While a lower number of clusters yields a reduced feature space and more interpretable outcomes, CAPs may not be segmented finely enough to resolve insightful dynamic properties regarding cognition. Our strategy was thus to first perform an exploratory assessment, in which we evaluated the significance of the brain/behaviour correlation across a set of candidate $K$ values ($K_{opt}=\{14,16,22,32\}$) and assignment thresholds (forcing the assignment of all frames, or using $T_\text{P}=[0:5:100]$), to select the relevant parameters to proceed forward with.

The results of this exploratory assessment are displayed in Figure~\ref{CHAPTER5_P1F3}B, when scrubbed frames are discarded (left panel) or also assigned to the CAPs at threshold $T_\text{P}$. Both settings yield very similar significance values, which is good evidence that remaining head motion effects only have a minor influence on the analyses. As the assignment threshold increases (that is, less and less frames are assigned because the criterion becomes more and more stringent), significance generally decreases. A smooth spot can be observed for $K=\{16,22\}$ and $T_P < 15$\%, which indicates that this granularity is optimal in the context of behavioural prediction. We selected $K=16$ and $T_P=0$\% as values for more detailed subsequent analyses.

\begin{figure}[p]
\centering
\includegraphics[width=1.0\textwidth]{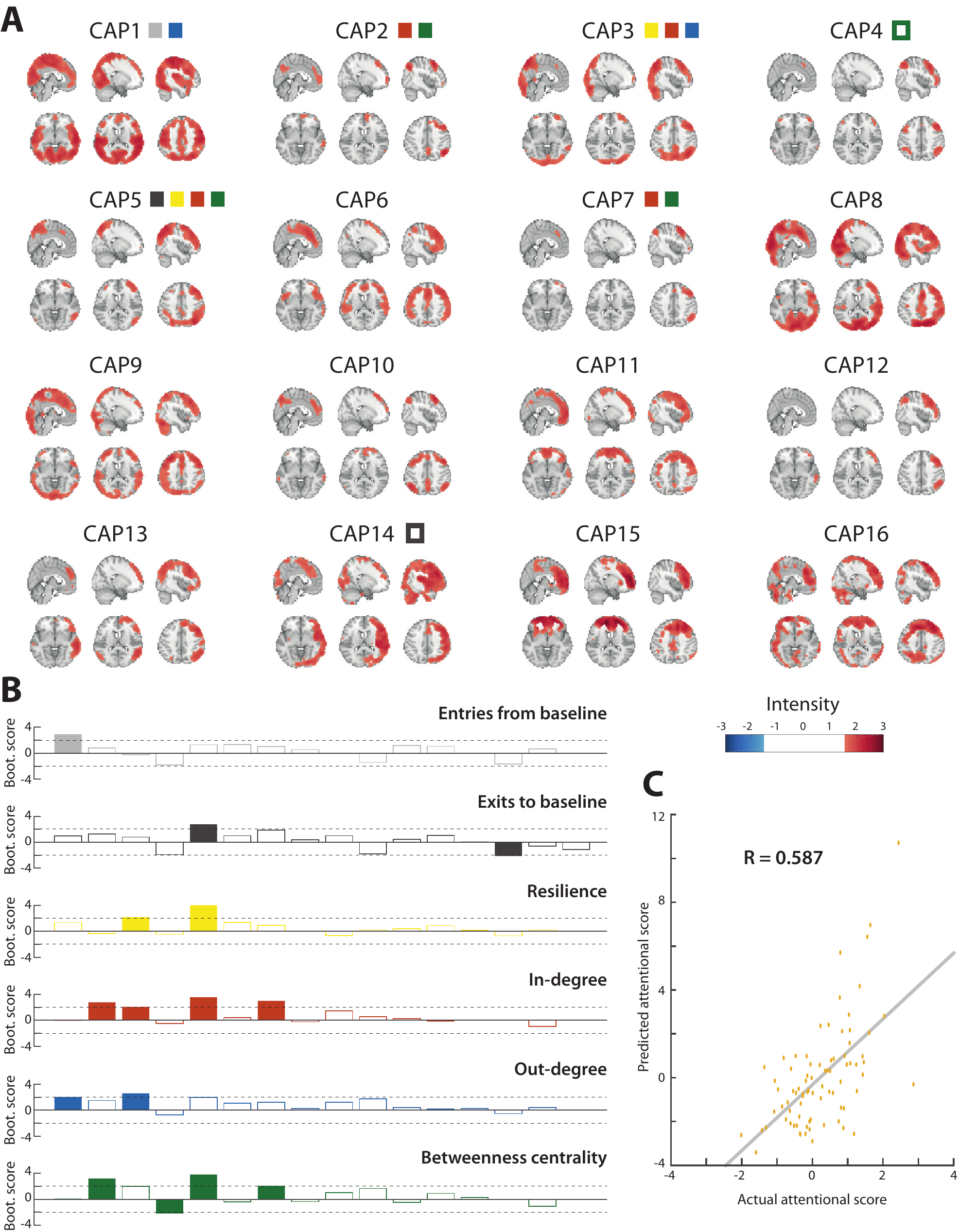}
\caption{\textbf{Co-activation patterns, and relationship to attentional abilities.} \textbf{(A)} The 16 obtained CAPs are plotted, with coloured rectangles symbolising the CAPs that are particularly important (bootstrap score of the associated salience weight larger than 2) for the brain/behaviour relationship as quantified from a given metric (grey: entries from baseline state, black: exits to baseline state, yellow: resilience, red: in-degree, blue: out-degree, green: betweenness centrality). Filled rectangles highlight beneficial CAPs, while hollow rectangles depict detrimental CAPs (negative bootstrap score). \textbf{(B)} Salience weights across all 16 CAPs and the 6 investigated CAP metrics. The threshold bootstrap score past which a weight is considered significant is highlighted by a horizontal dashed line. Empty bars denote non-significant weights, while filled bars represent significant ones.  \textbf{(C)} Actual attentional performance score (x-axis) versus predicted values with PLS (y-axis).}
\label{CHAPTER5_P1F4}
\end{figure}

CAPs are displayed, for this chosen parameter set, in Figure~\ref{CHAPTER5_P1F4}A, while their involvement in driving the brain/behaviour relationship, as quantified by salience weights across our range of investigated metrics, is depicted in Figure~\ref{CHAPTER5_P1F4}B. The correlation between actual and predicted attentional performance was strongly significant ($R=0.587$, $p<0.001$; Figure~\ref{CHAPTER5_P1F4}C). The associated covariance component found by PLS analysis was significant at $p=0.003$. Note that this relationship is derived from only subjects that were not used to construct the CAPs.

CAP\textsubscript{1} depicts co-activation of a range of resting-state networks, including the auditory, somatomotor, visual and salience ones. Attentional performance was better in the subjects that transited more frequently from the baseline state of seed activity to this CAP. CAP\textsubscript{1} was also more often the entry point towards other CAPs in high performance subjects, as seen from a strongly positive out-degree salience weight.

Good subjects in terms of continuous performance also more often entered CAP\textsubscript{2} and CAP\textsubscript{7} from other states (as seen from positive in-degree salience weights), and these same 2 CAPs were also more influential in the transitory behaviour of CAP dynamics (since betweenness centrality salience weights also showed large positive values). In both CAPs, the seed region co-activates with a restricted set of areas including the right inferior parietal cortex (for both), the posterior cingulate cortex and medial prefrontal cortex (for CAP\textsubscript{2}), and the right anterior prefrontal cortex (for CAP\textsubscript{7}).

CAP\textsubscript{3} and CAP\textsubscript{5} were associated to good attentional abilities from the viewpoint of several metrics, which emphasises the importance of their expression: for CAP\textsubscript{3}, it involved resilience, in-degree and out-degree, while for CAP\textsubscript{5}, it was return to baseline, resilience, in-degree and betweenness centrality. Both CAPs include strong co-activation with the right inferior parietal cortex, and for CAP\textsubscript{3}, also with the left cerebellum lobule VI and a subpart of the occipital cortex.

CAP\textsubscript{4} and CAP\textsubscript{14} were the only states whose expression was detrimental for attentional performance, in terms of betweenness centrality for the former, and of return to baseline for the latter. CAP\textsubscript{4} displayed bilateral right superior cortex and anterior prefrontal cortex co-activation with the seed, while for CAP\textsubscript{14}, involved areas were the fusiform gyrus, parahippocampal cortex, and a diffuse right lateralised spot covering parts of the auditory, secondary somatosensory and posterior insular cortices.

The majority of the other CAPs that did not show any link to attentional performance involved co-activations with regions that were not part of the attentional networks: for instance, CAP\textsubscript{6} includes the anterior cingulate cortex and anterior insula; CAP\textsubscript{8} contains the anterior cingulate, visual and right somatosensory cortices; CAP\textsubscript{9} showcases primary visual, and auditory cortices; CAP\textsubscript{10} shows the angular gyrus and part of the precuneus; CAP\textsubscript{11} and CAP\textsubscript{15} mostly highlight ventral medial prefrontal cortex signal.

\section{Discussion}
\label{sec:discussion}

In this work, we have introduced TbCAPs, a toolbox for co-activation pattern analysis, which provides practitioners with an intuitive dedicated graphical user interface as well as a powerful scripting equivalent. It provides an easy control over all key analytical parameters of the technique, novel methodological additions for augmented analyses, and facilitated visualisation of the resulting CAPs and associated metrics. Although we have focussed on the usefulness of CAP analysis in the resting-state setting, we also remark that nothing precludes the use of the technique in task-based investigations.

As most CAP studies to date have revealed the potential of the approach in clinical settings~\cite{Amico2014,DiPerri2018,Kaiser2019,Tuleasca2019}, we sought to demonstrate the relevance of the technique in another context; \textit{i.e.}, rather than considering a \textit{classification} problem in which two or more distinct subject populations are separated, we considered a \textit{regression} task in which we attempted to explain attentional abilities within a more homogeneous population in a continuous vigilance task by means of CAPs dynamics.

We observed that of all the extracted CAPs showing coupling with the right dorsolateral prefrontal seed, the large majority either did not appear to be involved in attentional abilities, or showed positive salience weights indicating a positive impact of their expression. This is not so surprising, given that our analysis was focused on a region of the attention network in the first place. The common feature of beneficial CAPs appeared to be the coupling of an array of other regions previously pinpointed in continuous performance tasks, including the inferior parietal cortex, cerebellum lobule VI or occipital cortex~\cite{Hager1998,Ogg2008,Tana2010}. At the same time, these beneficial CAPs also barely involved coupling of other functionally distinct networks.

The one CAP for which the above reasoning does not hold is CAP\textsubscript{1}: despite the involvement of a very diverse set of regions, it was also retrieved as beneficial for attentional performance. More precisely, contrarily to most of the others, the expression of this CAP appears to be essential at the start of a seed activation sequence: indeed, salience weights were large specifically for the entries from baseline, and out-degree metrics. In other words, there is first a transition from baseline to this CAP, followed by the exit of that configuration to reach more spatially well-defined states. This involvement of short-lived periods of extensive cross-network interactions in mediating some aspects of human cognition has recently started to be appreciated as an insightful functional brain mechanism~\cite{Betzel2016,Fukushima2018b}.

The fact that all the probed metrics significantly contributed to explain attention is good evidence in favour of the temporal complexity of functional brain dynamics: instead of an instantaneous characterisation or a one-frame expression of a telling functional state, what truly matters is a complex mix between how activation starts (captured by the from-baseline and to-baseline metrics), how transitions occur across distinct functional states (as seen from in-degree, out-degree and betweenness centrality), and how lasting a given state is (as quantified by resilience). Our characterisation can be placed in the broad family of temporal modelling approaches, of which other notable examples include the use of hidden Markov models~\cite{Vidaurre2017,Bolton2017b} or graph-theoretical analysis for energy landscape~\cite{Kang2019}.

Prediction of continuous performance abilities from resting-state fMRI recordings has been shown possible in previous functional connectivity work relying on second-order correlational measures across brain regions~\cite{Rosenberg2016}. More recently, this characterisation has been pushed to the dynamic level by~\cite{Fong2019}, who showed that prediction can also be successfully achieved when \textit{temporal variability}, which quantifies fluctuations in functional connectivity over the course of a scanning session, is used as a metric of interest. Our prediction accuracy is on par with the one achieved in this whole-brain analysis, despite focussing on one seed region. Interestingly, the authors described the fact that lowered temporal variability was beneficial for better attentional performance, and that many of the most important features for prediction involved the executive-control brain networks: this is fully consistent with our results, in which increased resilience (which can be expected to yield lowered temporal variability) of CAPs featuring executive-control areas is beneficial.

In comparison to clinical applications of CAP analysis, in which between $3$ to $8$ CAPs are typically considered, a finer granularity was required in the present work (see Figure~\ref{CHAPTER5_P1F3}). This is not surprising given that the regression problem at hand here is more challenging than a classification task, as we need to predict a value within a continuum. Furthermore, the functional underpinnings of inter-individual differences in cognitive abilities are likely more subtle than when comparing subjects across consciousness levels or disease severity levels. In fact, if a too low number of CAPs is extracted, patterns with a different cognitive relevance are averaged together as a single cluster, which impedes prediction; conversely, if a too large number of CAPs is extracted, meaningful configurations become further segmented, and statistical power is lost due to the smaller amount of frames constituting each CAP.

In future work, it will be interesting to examine clinical or cognitive research hypotheses at the broader focus level of more than one seed region. As alluded to above, this is already feasible with our current toolbox version, and may enable to better bridge the gap between seed-based and whole-brain analyses. We also foresee additional technical developments in the near future, such as the possibility to extract \textit{co-activation sequences} (that is, series of successive fMRI volumes) rather than CAPs.

\section{Acknowledgments}

Constantin Tuleasca gratefully acknowledges receipt of a 'Young Researcher in Clinical Research Grant' (Jeune Chercheur en Recherche Clinique) from the University of Lausanne (UNIL), Faculty of Biology and Medicine (FBM), and the Lausanne University Hospital (CHUV). In addition, the authors would like to thank Rapha\"{e}l Li\'{e}geois for his rereading of the manuscript.

\bibliographystyle{unsrt}
\bibliography{papers_library.bib}

\begin{thebibliography}{10}

\bibitem{Logothetis2001}
N~K Logothetis, J~Pauls, M~Augath, T~Trinath, and A~Oeltermann.
\newblock {Neurophysiological investigation of the basis of the fMRI signal}.
\newblock {\em Nature}, 412(6843):150--157, 2001.

\bibitem{Friston1994}
Karl~J Friston, Andrew~P Holmes, Keith~J Worsley, J-P Poline, Chris~D Frith,
  and Richard~SJ Frackowiak.
\newblock Statistical parametric maps in functional imaging: a general linear
  approach.
\newblock {\em Human brain mapping}, 2(4):189--210, 1994.

\bibitem{Friston1994b}
K.J. Friston.
\newblock {F}unctional and effective connectivity in neuroimaging: {A}
  synthesis.
\newblock {\em Human Brain Mapping}, 2(1-2):56--78, 1994.

\bibitem{Biswal1995}
Bharat Biswal, F~Zerrin~Yetkin, Victor~M Haughton, and James~S Hyde.
\newblock Functional connectivity in the motor cortex of resting human brain
  using echo-planar {MRI}.
\newblock {\em Magnetic resonance in medicine}, 34(4):537--541, 1995.

\bibitem{Fox2005}
Michael~D Fox, Abraham~Z Snyder, Justin~L Vincent, Maurizio Corbetta, David~C
  {Van Essen}, and Marcus~E Raichle.
\newblock {The human brain is intrinsically organized into dynamic,
  anticorrelated functional networks.}
\newblock {\em Proceedings of the National Academy of Sciences of the United
  States of America}, 102(27):9673--8, 2005.

\bibitem{Damoiseaux2006}
J~S Damoiseaux, S~A R~B Rombouts, F~Barkhof, P~Scheltens, C~J Stam, S~M Smith,
  and C~F Beckmann.
\newblock {Consistent resting-state networks across healthy subjects.}
\newblock {\em Proceedings of the National Academy of Sciences of the United
  States of America}, 103(37):13848--53, 2006.

\bibitem{Power2011}
Jonathan~D Power, Alexander~L Cohen, Steven~M Nelson, Gagan~S Wig, Kelly~Anne
  Barnes, Jessica~A Church, Alecia~C Vogel, Timothy~O Laumann, Fran~M Miezin,
  Bradley~L Schlaggar, et~al.
\newblock Functional network organization of the human brain.
\newblock {\em Neuron}, 72(4):665--678, 2011.

\bibitem{VanDenHeuvel2010}
Martijn~P. van~den Heuvel and Hilleke~E. {Hulshoff Pol}.
\newblock {Exploring the brain network: A review on resting-state fMRI
  functional connectivity}.
\newblock {\em European Neuropsychopharmacology}, 20(8):519--534, 2010.

\bibitem{Greicius2008}
Michael Greicius.
\newblock Resting-state functional connectivity in neuropsychiatric disorders.
\newblock {\em Current opinion in neurology}, 21(4):424--430, 2008.

\bibitem{Fox2010}
M.D. Fox and M.~Greicius.
\newblock {C}linical applications of resting state functional connectivity.
\newblock {\em Frontiers in Systems Neuroscience}, 4:19, 2010.

\bibitem{Chang2010}
Catie Chang and Gary~H. Glover.
\newblock {Time-frequency dynamics of resting-state brain connectivity measured
  with fMRI}.
\newblock {\em NeuroImage}, 50(1):81--98, 2010.

\bibitem{Hutchison2013b}
R.~Matthew Hutchison, Thilo Womelsdorf, Elena~A. Allen, Peter~A. Bandettini,
  Vince~D. Calhoun, Maurizio Corbetta, Stefania {Della Penna}, Jeff~H. Duyn,
  Gary~H. Glover, Javier Gonzalez-Castillo, Daniel~A. Handwerker, Shella
  Keilholz, Vesa Kiviniemi, David~A. Leopold, Francesco de~Pasquale, Olaf
  Sporns, Martin Walter, and Catie Chang.
\newblock {Dynamic functional connectivity: Promise, issues, and
  interpretations}.
\newblock {\em NeuroImage}, 80:360--378, 2013.

\bibitem{Preti2017}
Maria~Giulia Preti, Thomas~AW Bolton, and Dimitri {Van De Ville}.
\newblock {The dynamic functional connectome: State-of-the-art and
  perspectives}.
\newblock {\em NeuroImage}, 160(December 2016):41--54, 2017.

\bibitem{Tagliazucchi2012b}
Enzo Tagliazucchi, Pablo Balenzuela, Daniel Fraiman, and Dante~R Chialvo.
\newblock {Criticality in Large-Scale Brain fMRI Dynamics Unveiled by a Novel
  Point Process Analysis}.
\newblock {\em Frontiers in Physiology}, 3(February):1--12, 2012.

\bibitem{VanEssen2013}
David~C. {Van Essen}, Stephen~M. Smith, Deanna~M. Barch, Timothy~E.J. Behrens,
  Essa Yacoub, and Kamil Ugurbil.
\newblock {The WU-Minn Human Connectome Project: An overview}.
\newblock {\em NeuroImage}, 80:62--79, 2013.

\bibitem{Nooner2012}
Kate~Brody Nooner, Stanley Colcombe, Russell Tobe, Maarten Mennes, Melissa
  Benedict, Alexis Moreno, Laura Panek, Shaquanna Brown, Stephen Zavitz,
  Qingyang Li, et~al.
\newblock The {NKI}-{R}ockland sample: a model for accelerating the pace of
  discovery science in psychiatry.
\newblock {\em Frontiers in neuroscience}, 6:152, 2012.

\bibitem{Holmes2015}
Avram~J Holmes, Marisa~O Hollinshead, Timothy~M O'Keefe, Victor~I Petrov,
  Gabriele~R Fariello, Lawrence~L Wald, Bruce Fischl, Bruce~R Rosen, Ross~W
  Mair, Joshua~L Roffman, et~al.
\newblock Brain genomics superstruct project initial data release with
  structural, functional, and behavioral measures.
\newblock {\em Scientific data}, 2, 2015.

\bibitem{Liu2013}
Xiao Liu and Jeff~H Duyn.
\newblock {Time-varying functional network information extracted from brief
  instances of spontaneous brain activity.}
\newblock {\em Proceedings of the National Academy of Sciences of the United
  States of America}, 110(11):4392--7, 2013.

\bibitem{Amico2014}
Enrico Amico, Francisco Gomez, Carol {Di Perri}, Audrey Vanhaudenhuyse, Damien
  Lesenfants, Pierre Boveroux, Vincent Bonhomme, Jean~Fran{\c{c}}ois Brichant,
  Daniele Marinazzo, and Steven Laureys.
\newblock {Posterior cingulate cortex-related co-activation patterns: A resting
  state fMRI study in propofol-induced loss of consciousness}.
\newblock {\em PLoS ONE}, 9(6):1--9, 2014.

\bibitem{DiPerri2018}
Carol Di~Perri, Enrico Amico, Lizette Heine, Jitka Annen, Charlotte Martial,
  Stephen~Karl Larroque, Andrea Soddu, Daniele Marinazzo, and Steven Laureys.
\newblock Multifaceted brain networks reconfiguration in disorders of
  consciousness uncovered by co-activation patterns.
\newblock {\em Human brain mapping}, 39(1):89--103, 2018.

\bibitem{Kaiser2019}
Roselinde~H Kaiser, Min~Su Kang, Yechan Lew, Julie Van Der~Feen, Blaise
  Aguirre, Rachel Clegg, Franziska Goer, Erika Esposito, Randy~P Auerbach,
  R~Matthew Hutchison, et~al.
\newblock Abnormal frontoinsular-default network dynamics in adolescent
  depression and rumination: a preliminary resting-state co-activation pattern
  analysis.
\newblock {\em Neuropsychopharmacology}, page~1, 2019.

\bibitem{Tuleasca2019}
Constantin Tuleasca, Thomas~AW Bolton, Jean R{\'e}gis, Elena Najdenovska,
  Tatiana Witjas, Nadine Girard, Francois Delaire, Marion Vincent, Mohamed
  Faouzi, Jean-Philippe Thiran, et~al.
\newblock Normalization of aberrant pretherapeutic dynamic functional
  connectivity of extrastriate visual system in patients who underwent
  thalamotomy with stereotactic radiosurgery for essential tremor: a
  resting-state functional mri study.
\newblock {\em Journal of neurosurgery}, 1(aop):1--10, 2019.

\bibitem{Di2013}
Xin Di and Bharat~B. Biswal.
\newblock {Dynamic brain functional connectivity modulated by resting-state
  networks}.
\newblock {\em Brain Structure and Function}, 220(1):37--46, jan 2015.

\bibitem{Liu2013b}
Xiao Liu, Catie Chang, and Jeff~H Duyn.
\newblock {Decomposition of spontaneous brain activity into distinct fMRI
  co-activation patterns}.
\newblock {\em Frontiers in Systems Neuroscience}, 7(December):1--11, 2013.

\bibitem{Chen2015}
Jingyuan~E. Chen, Catie Chang, Michael~D. Greicius, and Gary~H. Glover.
\newblock {Introducing co-activation pattern metrics to quantify spontaneous
  brain network dynamics}.
\newblock {\em NeuroImage}, 111:476--488, 2015.

\bibitem{Zhuang2018}
Xiaowei Zhuang, Ryan~R Walsh, Karthik Sreenivasan, Zhengshi Yang, Virendra
  Mishra, and Dietmar Cordes.
\newblock Incorporating spatial constraint in co-activation pattern analysis to
  explore the dynamics of resting-state networks: An application to
  {P}arkinson's disease.
\newblock {\em NeuroImage}, 172:64--84, 2018.

\bibitem{Liu2018}
Xiao Liu, Nanyin Zhang, Catie Chang, and Jeff~H. Duyn.
\newblock {Co-activation patterns in resting-state fMRI signals}.
\newblock {\em NeuroImage}, 180(September 2017):485--494, 2018.

\bibitem{Power2012}
Jonathan~D. Power, Kelly~A. Barnes, Abraham~Z. Snyder, Bradley~L. Schlaggar,
  and Steven~E. Petersen.
\newblock Spurious but systematic correlations in functional connectivity {MRI}
  networks arise from subject motion.
\newblock {\em NeuroImage}, 59(3):2142--2154, 2012.

\bibitem{VanDijk2012}
K.R. Van~Dijk, M.R. Sabuncu, and R.L. Buckner.
\newblock {T}he influence of head motion on intrinsic functional connectivity
  {MRI}.
\newblock {\em NeuroImage}, 59(1):431--438, 2012.

\bibitem{Satterthwaite2012}
Theodore~D Satterthwaite, Daniel~H Wolf, James Loughead, Kosha Ruparel, Mark~A
  Elliott, Hakon Hakonarson, Ruben~C Gur, and Raquel~E Gur.
\newblock {Impact of in-scanner head motion on multiple measures of functional
  connectivity : Relevance for studies of neurodevelopment in youth}.
\newblock {\em NeuroImage}, 60(1):623--632, 2012.

\bibitem{Shirer2012}
W.~R. Shirer, S.~Ryali, E.~Rykhlevskaia, V.~Menon, and M.~D. Greicius.
\newblock {Decoding subject-driven cognitive states with whole-brain
  connectivity patterns}.
\newblock {\em Cerebral Cortex}, 22(1):158--165, 2012.

\bibitem{Monti2003}
S.~Monti, P.~Tamayo, J.~Mesirov, and T.~Golub.
\newblock {C}onsensus clustering: a resampling-based method for class discovery
  and visualization of gene expression microarray data.
\newblock {\em Machine Learning}, 52(1-2):91--118, 2003.

\bibitem{Rubinov2010}
Mikail Rubinov and Olaf Sporns.
\newblock {Complex network measures of brain connectivity: Uses and
  interpretations}.
\newblock {\em NeuroImage}, 52(3):1059--1069, 2010.

\bibitem{Freeman1979}
Linton~C Freeman.
\newblock Centrality in social networks conceptual clarification.
\newblock {\em Social networks}, 1(3):215--239, 1978.

\bibitem{Senbabaoglu2014}
Yasin Senbabaoglu, George Michailidis, and Jun~Z Li.
\newblock {Critical limitations of consensus clustering in class discovery}.
\newblock {\em Scientific Reports}, 4(6207), 2014.

\bibitem{Smith2013}
Stephen~M Smith, Christian~F Beckmann, Jesper Andersson, Edward~J Auerbach,
  Janine Bijsterbosch, Gwena{\"e}lle Douaud, Eugene Duff, David~A Feinberg,
  Ludovica Griffanti, Michael~P Harms, et~al.
\newblock Resting-state f{MRI} in the human connectome project.
\newblock {\em Neuroimage}, 80:144--168, 2013.

\bibitem{Bolton2019c}
Thomas~AW Bolton, Daniela Z{\"o}ller, C{\'e}sar Caballero-Gaudes, Valeria
  Kebets, Enrico Glerean, and Dimitri Van De~Ville.
\newblock Agito ergo sum: correlates of spatiotemporal motion characteristics
  during f{MRI}.
\newblock {\em arXiv preprint arXiv:1906.06445}, 2019.

\bibitem{Murphy2017}
Kevin Murphy and Michael~D Fox.
\newblock Towards a consensus regarding global signal regression for resting
  state functional connectivity {MRI}.
\newblock {\em Neuroimage}, 154:169--173, 2017.

\bibitem{Gur2010}
Ruben~C Gur, Jan Richard, Paul Hughett, Monica~E Calkins, Larry Macy, Warren~B
  Bilker, Colleen Brensinger, and Raquel~E Gur.
\newblock A cognitive neuroscience-based computerized battery for efficient
  measurement of individual differences: standardization and initial construct
  validation.
\newblock {\em Journal of neuroscience methods}, 187(2):254--262, 2010.

\bibitem{Bishop1999}
C.M. Bishop.
\newblock {B}ayesian {PCA}, 1999.

\bibitem{Kriegeskorte2009}
N~Kriegeskorte, W.~K. Simmons, P.~S.~F. Bellgowan, and C.~I. Baker.
\newblock Circular analysis in systems neuroscience: the dangers of double
  dipping.
\newblock {\em Nature Neuroscience}, 12:535--540, 2009.

\bibitem{McIntosh2004}
A.~R. McIntosh and N.~J. Lobaugh.
\newblock Partial least squares analysis of neuroimaging data: applications and
  advances.
\newblock {\em NeuroImage}, 23, Supplement 1:S250--S263, 2004.
\newblock Mathematics in Brain Imaging.

\bibitem{Krishnan2011}
A.~Krishnan, L.~J. Williams, A.~R. McIntosh, and H.~Abdi.
\newblock Partial least squares {(PLS)} methods for neuroimaging: A tutorial
  and review.
\newblock {\em NeuroImage}, 56(2):455--475, 2011.
\newblock Multivariate Decoding and Brain Reading.

\bibitem{Hager1998}
Frank H{\"a}ger, Hans-Peter Volz, Christian Gaser, Hans-Joachim Mentzel,
  Werner~A Kaiser, and Heinrich Sauer.
\newblock Challenging the anterior attentional system with a continuous
  performance task: a functional magnetic resonance imaging approach.
\newblock {\em European Archives of Psychiatry and Clinical Neuroscience},
  248(4):161--170, 1998.

\bibitem{Ogg2008}
Robert~J Ogg, Ping Zou, Deanna~N Allen, Sabrina~B Hutchins, Radek~M Dutkiewicz,
  and Raymond~K Mulhern.
\newblock Neural correlates of a clinical continuous performance test.
\newblock {\em Magnetic resonance imaging}, 26(4):504--512, 2008.

\bibitem{Tana2010}
Maria~Gabriella Tana, Eros Montin, Sergio Cerutti, and Anna~M Bianchi.
\newblock Exploring cortical attentional system by using f{MRI} during a
  continuous perfomance test.
\newblock {\em Computational intelligence and neuroscience}, 2010:3, 2010.

\bibitem{Betzel2016}
Richard~F Betzel, Makoto Fukushima, Ye~He, Xi-nian Zuo, and Olaf Sporns.
\newblock {Dynamic fluctuations coincide with periods of high and low
  modularity in resting-state functional brain networks}.
\newblock {\em NeuroImage}, 127:287--297, feb 2016.

\bibitem{Fukushima2018b}
Makoto Fukushima, Richard~F Betzel, Ye~He, Marcel~A de~Reus, Martijn~P van~den
  Heuvel, Xi-Nian Zuo, and Olaf Sporns.
\newblock Fluctuations between high-and low-modularity topology in
  time-resolved functional connectivity.
\newblock {\em NeuroImage}, 180:406--416, 2018.

\bibitem{Vidaurre2017}
Diego Vidaurre, Stephen~M. Smith, and Mark~W. Woolrich.
\newblock {Brain network dynamics are hierarchically organized in time}.
\newblock {\em Proceedings of the National Academy of Sciences},
  114(48):201705120, 2017.

\bibitem{Bolton2017b}
Thomas~AW Bolton, Anjali Tarun, Virginie Sterpenich, Sophie Schwartz, and
  Dimitri Van De~Ville.
\newblock Interactions between large-scale functional brain networks are
  captured by sparse coupled {HMM}s.
\newblock {\em IEEE transactions on medical imaging}, 37(1):230--240, 2017.

\bibitem{Kang2019}
Jiyoung Kang, Chongwon Pae, and Hae-Jeong Park.
\newblock Graph-theoretical analysis for energy landscape reveals the
  organization of state transitions in the resting-state human cerebral cortex.
\newblock {\em PloS one}, 14(9):e0222161, 2019.

\bibitem{Rosenberg2016}
M.D. Rosenberg, E.S. Finn, D.~Scheinost, X.~Papademetris, X.~Shen, R.T.
  Constable, and M.M. Chun.
\newblock {A} neuromarker of sustained attention from whole-brain functional
  connectivity.
\newblock {\em Nature Neuroscience}, 19(1):165, 2016.

\bibitem{Fong2019}
Angus Ho~Ching Fong, Kwangsun Yoo, Monica~D Rosenberg, Sheng Zhang,
  Chiang-Shan~R Li, Dustin Scheinost, R~Todd Constable, and Marvin~M Chun.
\newblock Dynamic functional connectivity during task performance and rest
  predicts individual differences in attention across studies.
\newblock {\em NeuroImage}, 188:14--25, 2019.

\end{thebibliography}

\end{document}